\documentclass[aip,graphicx]{revtex4-1}
\usepackage{siunitx}
\usepackage{chemformula}
\usepackage{epstopdf}
\draft 

\begin{document}


\title{\ch{Si3N4} membrane as entrance window for plasma-generated vacuum ultraviolet (VUV) radiation} 



\author{Luka Hansen}
\email[]{lhansen@physik.uni-kiel.de}
\affiliation{Institute of Experimental and Applied Physics, Kiel University, Kiel, Germany}
\affiliation{Kiel Nano, Surface and Interface Science KiNSIS, Kiel University, Kiel, Germany}

\author{Görkem Bilgin}
\affiliation{Institute of Experimental and Applied Physics, Kiel University, Kiel, Germany}
\author{Jan Benedikt}
\affiliation{Institute of Experimental and Applied Physics, Kiel University, Kiel, Germany}
\affiliation{Kiel Nano, Surface and Interface Science KiNSIS, Kiel University, Kiel, Germany}



\date{\today}

\begin{abstract}
Vacuum ultraviolet (VUV) radiation produced by an atmospheric pressure plasma was successfully measured down to wavelengths of \SI{58.4}{nm} utilizing a \SI{20}{nm} thin \ch{Si3N4} membrane to transfer the VUV radiation into a vacuum monochromator.
This method allows measurements without disturbing the plasma or the spectra.
\ch{He2} absorption could be observed by filling the monochromator with \ch{He}.
Transmission of the \ch{Si3N4} membrane in the region of the \ch{He2}$^*$ excimer continua (\SIrange{58}{100}{nm}) could indirectly be measured and confirms literature values.
\end{abstract}

\pacs{}

\maketitle 


Vacuum ultraviolet (VUV) radiation corresponds to the \SIrange{10}{200}{nm} (\SIrange{124}{6.2}{eV}) wavelength range and has several applications ranging from biomedical over material modification to catalysis~\cite{Sze2020,Zos2014,Bak2019,Wan2023,Sun2023}.
As the name already suggests, the high energy photons have short lifetimes under atmospheric pressure conditions due to absorption from ambient species~\cite{Kel2013,Hea2017}.
Therefore, their use or measurement under atmospheric pressure conditions is quite challenging.
For diagnostics, the transfer of VUV photons produced under atmospheric pressure conditions into a vacuum environment is necessary to avoid absorption.
Commonly used window materials like \ch{LiF} or  \ch{MgF2} are not suited for this transfer, due to their cutoff wavelength at around \SI{110}{nm}~\cite{Ger2006}.
Therefore, different approaches have been pursuit in the past by omitting a window by either using differential pumping systems~\cite{Kur2001,Liu2020,Fie2021} or an aerodynamic window~\cite{Gol2020} to guide the VUV photons through a pure helium atmosphere, which was assumed to be transparent for VUV photons in the wavelength region longer than \SI{50.48}{nm} except for absorption at the He resonance line at \SI{58.43}{nm}~\cite{Lee1955,Bur1965,Mar1976,Sam1976, Hen1993}, neglecting \ch{He2} absorption~\cite{Tan1969,Cho1971,San1971,San1972}.\\
A new, window-based approach is presented in this work by using a \SI{20}{nm} thin \ch{Si3N4} membrane as entrance window to overcome the absorption inside of the window material by its low thickness.
The transmission of a \SI{20}{nm} thick \ch{Si3N4} membrane ranges from \SIrange{5}{14}{\%} within the wavelength region of interest in this study (\SIrange{120}{55}{nm}) and increases further at shorter wavelength~\cite{Hen1993,CXRO2024}.

\begin{figure}
    \centering
        \includegraphics[width=0.5\textwidth]{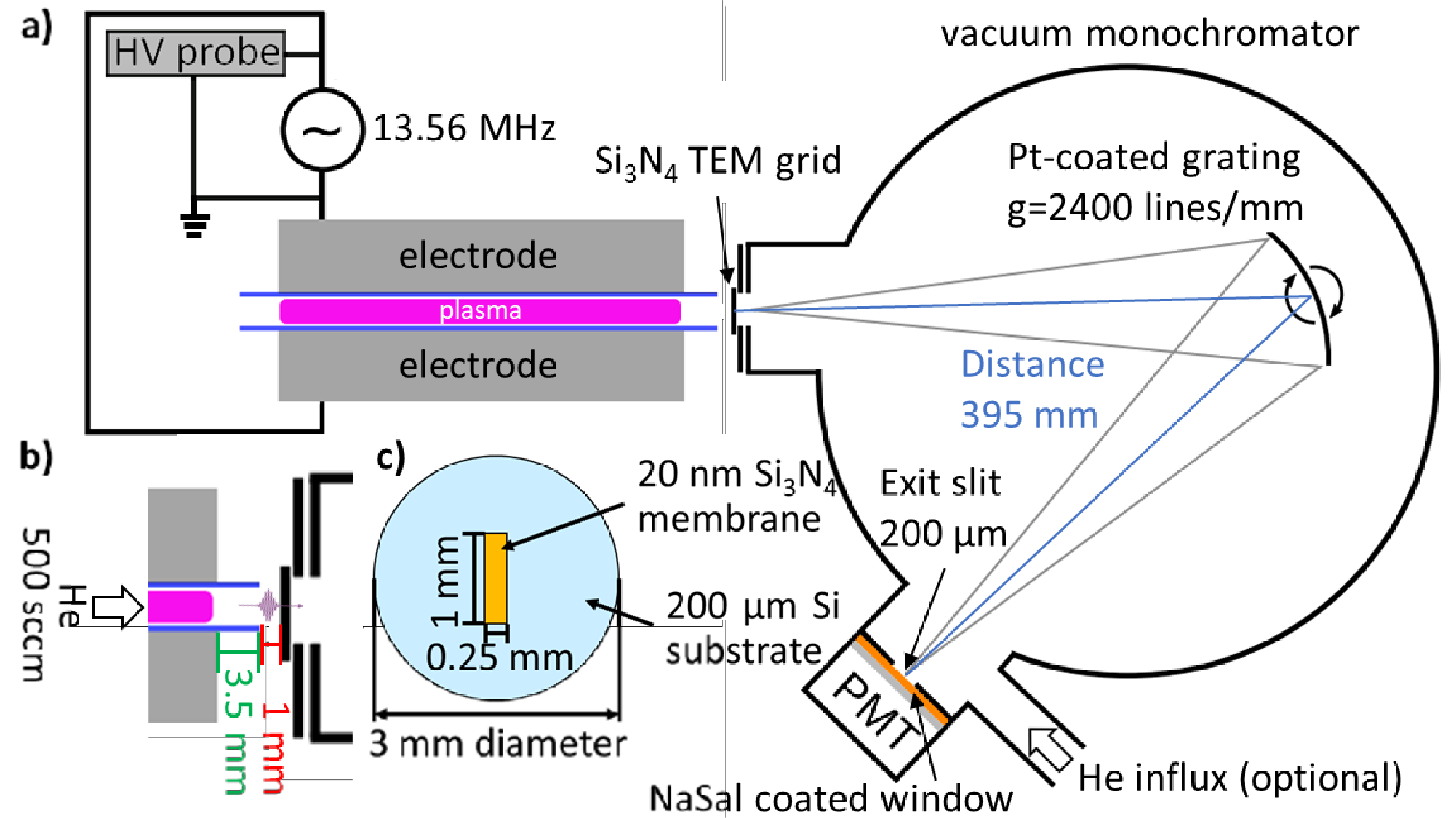}
    \caption{Schematic sketch of the experimental setup. a) Overview of the plasma source and the vacuum monochromator. b) Magnification of the entrance area of the monochromator. c) Sketch of the \ch{Si3N4} membrane used as entrance window.}
    \label{fig:setup}
\end{figure}

\begin{figure}
    \centering
        \includegraphics[width=0.45\textwidth]{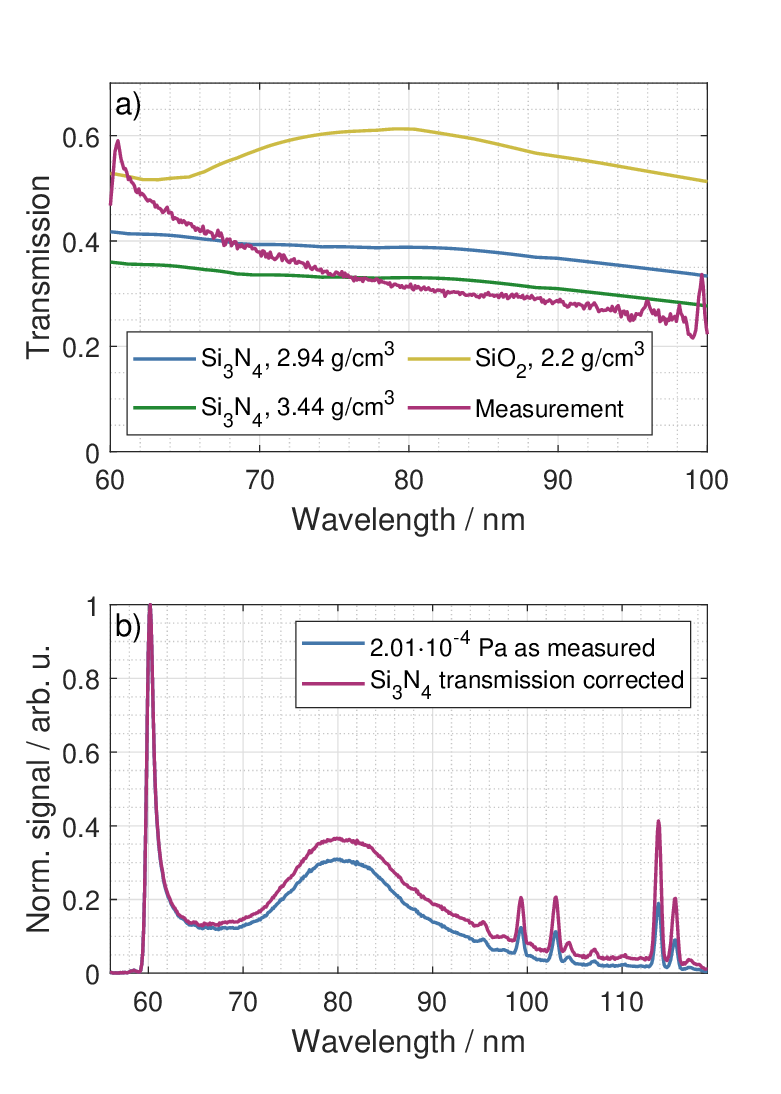}
    \caption{a) Transmission through \SI{10}{nm} thick \ch{Si3N4} or \ch{SiO2} membranes with different densities as given by literature~\cite{Hen1993,CXRO2024} compared to experimental determined transmission of a \SI{10}{nm} \ch{Si3N4} membrane equivalent as explained in the text. b) VUV spectra normalized to the maximum of the first \ch{He2}$^*$ excimer continuum and rescaled taking the transmission of the \ch{Si3N4} membrane according to~\cite{Hen1993,CXRO2024} into account.}
    \label{fig:rescaledSpectra}
\end{figure}

As VUV photon source, an atmospheric pressure plasma jet, the so-called capillary jet~\cite{Win2022}, was chosen, as the \ch{Si3N4} membranes showed resistance against plasma exposure~\cite{Tai2013,Han2023} and atmospheric pressure (micro) plasmas proved to produce VUV radiation~\cite{Kur2001,Sat2014,Par2017,Liu2020,Gol2020}. 
The setup is shown in Fig.~\ref{fig:setup} and consists of a Seya-Namioka design vacuum monochromator (Minuteman 302-VM, McPherson) with a focal length of \SI{200}{mm}, which was equipped with a \SI{20}{nm} thick \ch{Si3N4} thin film coated on a \SI{3}{mm} diameter \ch{Si} substrate with a thickness of \SI{200}{\textup{µm}}. In the center the \ch{Si} substrate is etched away resulting in a free standing membrane with dimensions of \SI{0.25}{mm} x \SI{1}{mm} (S171-9H, agarscientific, commercially available), as shown in Fig.~\ref{fig:setup}c).
The \ch{Si3N4} membrane served directly as entrance slit.
A Pt-coated rotating concave diffraction grating (\SI{2400}{lines/mm}) focused the diffracted light onto a \SI{200}{\textup{µm}} wide exit slit.
The photons interact with a sodium salicylate (\ch{NaSal}) layer, which converts the incident VUV photons directly proportional into visible photons in the \SIrange{350}{550}{nm} region~\cite{Sam1980}, which are detected by a photo multiplier tube (PMT, H7711-12, Hamamatsu). The spectra shown in this work were taken with \SI{0.1}{nm} increment steps and \num{24000} measurements per step at a sample rate of \SI{30}{kHz}. A spectrum ranging from \SIrange{55}{150}{nm} therefore took roughly \SI{14}{mins} including the time necessary for repositioning the grating, but can be accelerated if necessary. 
Prior to the measurements, a deuterium lamp was used for the wavelength calibration of the monochromator.\\
The capillary jet is set up from a rounded-edge rectangular glass capillary (Rect. Boro Capillaries 2540-100, CM Scientific) with inner dimensions of \SI{0.4}{mm} x \SI{4}{mm} and a wall thickness of \SI{0.28}{mm} mounted in between two stainless steel electrodes with an area of \SI{4}{mm} x \SI{40}{mm} in contact with the capillary. 
The capillary edge surpasses the electrodes by \SI{3.5}{mm} and the exit of the capillary was mounted with a distance of \SI{1}{mm} to the \ch{Si3N4} membrane as shown in Fig.~\ref{fig:setup}b).\\
The plasma was generated by flowing \SI{500}{sccm} of \ch{He} (5.0, Air Liquide) through the capillary in the direction of the \ch{Si3N4} membrane and applying \SI{480}{V_{\mathrm{pp}}} (\SI{50}{W} generator power) to the electrodes using a combination of a \SI{13.56}{MHz} radiofrequency (RF) generator (RFG-100/13, Coaxial Power Systems) with a matchbox (MMN 300-13, Coaxial Power Systems). 
The applied voltage was monitored with a high voltage probe (P6015A, Tektronix) to ensure the stability of the plasma.\\
The monochromator was evacuated to a pressure of \SI{2.01e-4}{Pa} and stepwise filled with \ch{He} up to a slight overpressure of \SI{1.08e5}{Pa} set by an overpressure valve. A flow of \ch{He} through the monochromator was present during the measurements, which in combination with adjusting the pumping speed resulted in the stable pressures and low level of impurities during the actual measurements.\\
The \ch{Si3N4} membrane used for the presented study served as an entrance window for \SI{47}{days} with in total \SI{12}{h} of plasma treatment/irradiation before being removed for a transmission electron microscopy (TEM) analysis and comparison with an untreated, pristine \ch{Si3N4} membrane from the same manufacturing batch. Energy dispersive X-ray spectroscopy (EDX) showed increased oxygen content of the treated membrane (\ch{Si}/\ch{N}/\ch{O}/\ch{C} ratio in \si{at\%}; pristine: 31/39/7/23 vs. treated: 27/14/33/26; C content probably due to contamination), which seems to not affect the transmission (see Fig.~\ref{fig:rescaledSpectra}a) for comparison with \ch{Si3N4} and \ch{SiO2} transmission literature values) as the measurement for the transmission was done directly before removing the membrane. Also the mechanical stability of the membrane was unaffected by this oxidation, which will be addressed in more detail in future work. 

\begin{figure*}
    \centering
        \includegraphics[width=1\textwidth]{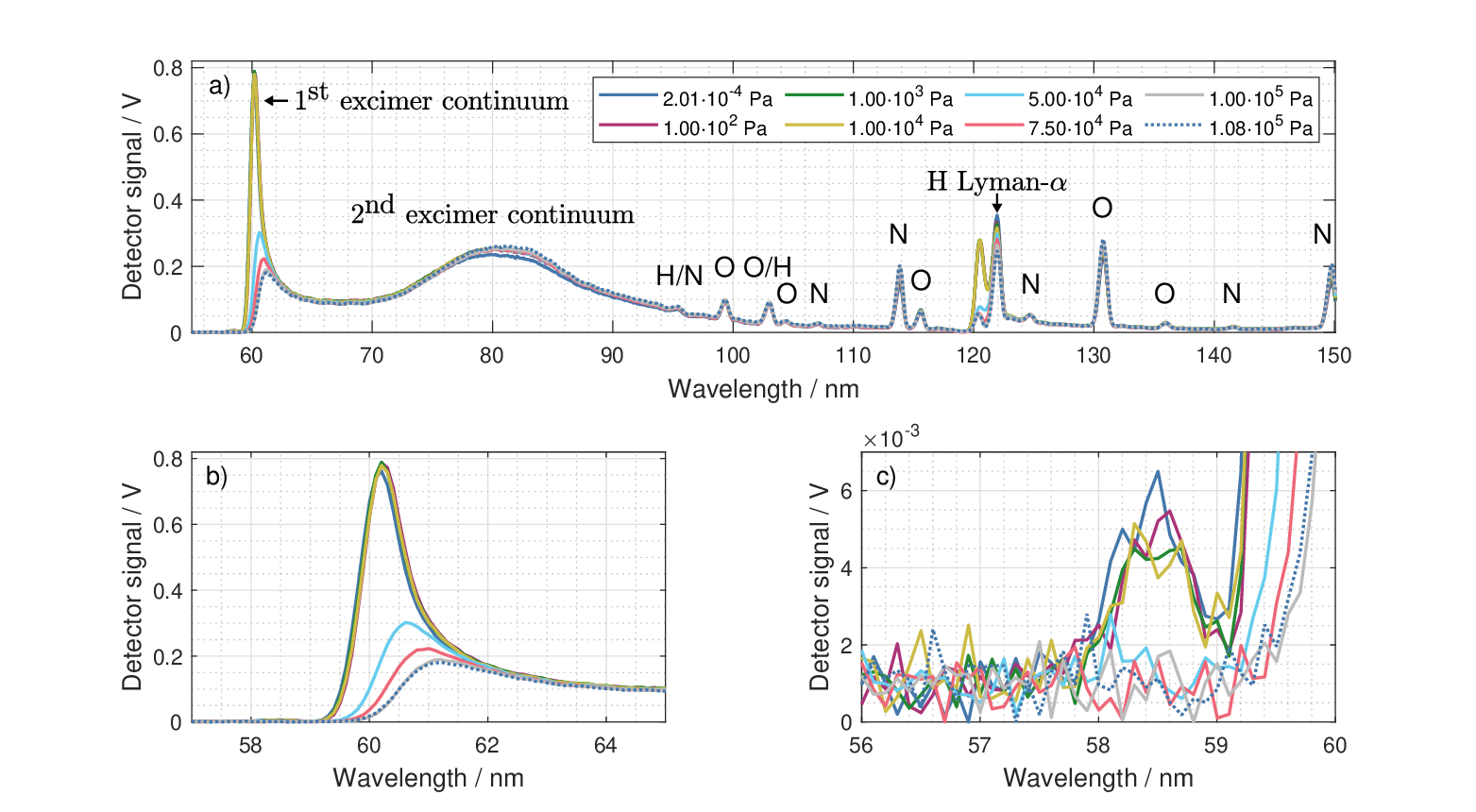}
    \caption{a) Measured VUV spectra in dependence of the \ch{He} pressure inside of the monochromator. b) Enlarged depiction of the first \ch{He2}$^*$ excimer continuum. c) Enlarged depiction of the \ch{He} resonance line.}
    \label{fig:spectra}
\end{figure*}

The results are shown in Fig.~\ref{fig:spectra} and show a clear dependence on the monochromator pressure in the region of the first \ch{He2}$^*$ excimer continuum ranging from \SIrange{58}{64}{nm}~\cite{Lym1924,Kur2001,Sgo2024} and also in the second \ch{He2}$^*$ excimer continuum ranging from \SIrange{64}{100}{nm}~\cite{Hop1930,Tan1958,Huf1962,Tan1962,Huf1965,Kur2001,Sgo2024}.
At \SI{120}{nm} the second diffraction order of the first continuum is visible.
Further visible atomic lines are annotated in Fig.~\ref{fig:spectra}a) and were identified using the NIST database~\cite{Kra1999}.
Fig.~\ref{fig:spectra}b) shows the first continuum and its changes starting at a pressure of \SI{5e4}{Pa}. 
The intensity is reduced by a factor of $\approx$~5 comparing the evacuated monochromator with the one under slight overpressure.
Further, the maximum position is shifted from \SI{60.1}{nm} to \SI{61.2}{nm} and the lower limit of the intensity onset is shifted from \SI{59.1}{nm} to \SI{59.8}{nm}.
Starting from the same pressure of \SI{5e4}{Pa}, also the \ch{He} resonance line at \SI{58.4}{nm}, shown in Fig.~\ref{fig:spectra}c) vanishes and gets absorbed by the \ch{He} atmosphere inside of the monochromator.

\begin{figure}
    \centering
        \includegraphics[width=0.45\textwidth]{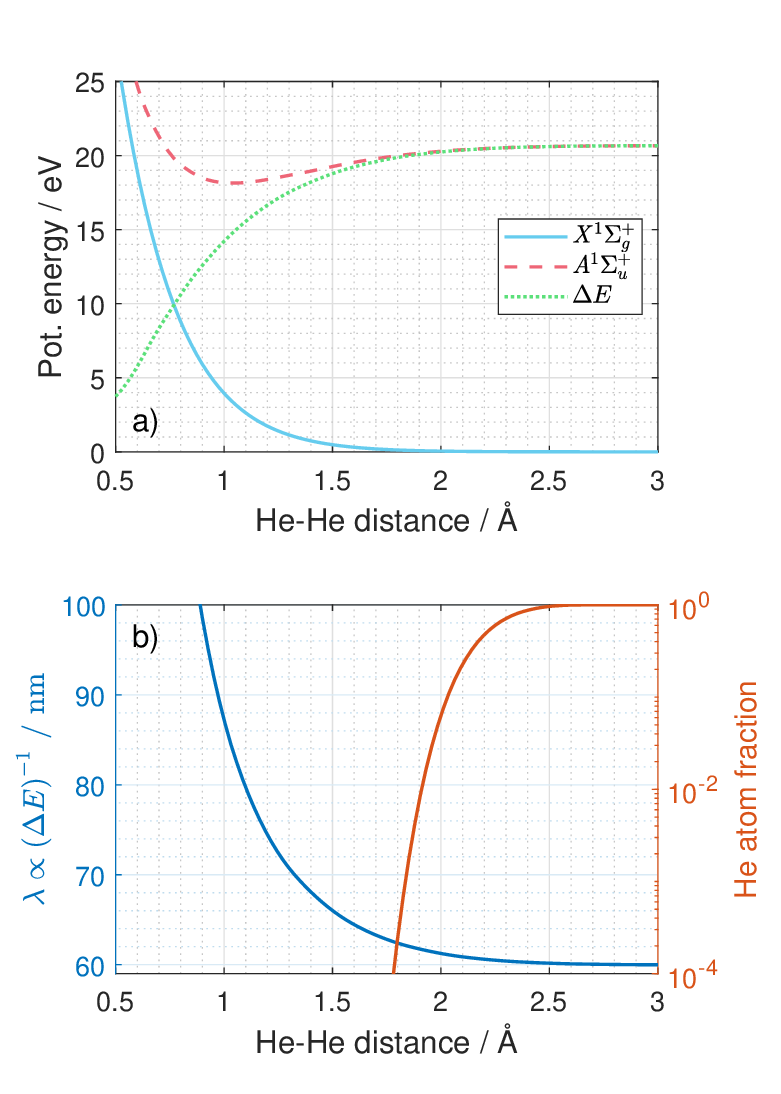}
    \caption{a) Potential energy curves for \ch{He2}($X$)~\cite{Prz2010} and \ch{He2}($A$)~\cite{Kom2006}, as well as their difference in potential energy. b) Wavelength corresponding to the potential energy difference between the \ch{He2}($X$) and \ch{He2}($A$) state (left, blue) and fraction of Maxwell-Boltzmann distributed \ch{He} atoms at \SI{295.15}{K} with the corresponding distance they can reach according to the potential energy of \ch{He2}($X$) (right, red).}
    \label{fig:PotEnergy}
\end{figure}

These measurement clearly showcase the results of \ch{He2} absorption~\cite{Tan1969,Cho1971,San1971,San1972} and the necessity of the window-based approach.
Utilizing an aerodynamic window~\cite{Gol2020} had the clear advantage of making VUV measurements from atmospheric pressure plasma in principle possible without influencing the plasma itself like it is the case for differential pumping systems~\cite{Kur2001,Liu2020,Fie2021}.
But neglecting the \ch{He2} absorption results in systematic errors in the region of the first \ch{He2}$^*$ excimer continuum and falsifies the ratio in between the two excimer continua.
The \ch{He2} absorption itself is due to photoinduced excimer formation inside of the monochromator.
A certain fraction of neutral \ch{He} atoms can reach internuclear distances in the non-binding $X^1 \sum_g^+$ state, which enables the transition into the metastable $A^1 \sum_u^+$ state.
The potential energy curves of the respective states are show in Fig.~\ref{fig:PotEnergy}a)~\cite{Kom2006,Prz2010}.
The \ch{He} fraction is illustrated in Fig.~\ref{fig:PotEnergy}b) on the right axis (red) based on a Maxwell-Boltzmann distribution for the relative kinetic energy ($E$),
\begin{equation}
f_E(E)=2*\sqrt{\frac{2E}{\pi}}*\left(\frac{1}{k_b*T}\right)^{3/2}*\text{exp}\left(\frac{-2E}{k_b*T}\right),    
\end{equation}
of the \ch{He} atoms inside of the monochromator with a temperature ($T$) of \SI{295.15}{K} (room temperature).
The potential energy curve of the non-binding $X^1 \sum_g^+$ state can be used to estimate, which fraction of the \ch{He} atoms can reach the given interatomic distance.
In the measurements effects of absorption are visible up to roughly \SI{62}{nm} which corresponds to a \ch{He} atom fraction of approx. \SI{0.2}{\%}.
The absorption is therefore classified as continuum-bound absorption~\cite{Cho1971} with a diffuse rotational fine structure~\cite{Tan1969}, which is not visible within the resolution of the used monochromator in its current state.

The influence of the membrane itself onto the VUV measurement of course has to be taken into account.
The transmission of the \ch{Si3N4} membrane is available in literature~\cite{Hen1993,CXRO2024} but depends on the \ch{Si3N4} density.
The \ch{Si3N4} density is typically \SI{3.44}{\g\per\cubic\cm}, but there is literature available reporting a density for similar membranes of \SI{2.94}{\g\per\cubic\cm}$\,$~\cite{She2019}.
Therefore, it is necessary to determine the transmission experimentally by performing VUV measurements under identical conditions beside an exchange of the \ch{Si3N4} membrane from \SI{20}{nm} thickness to \SI{30}{nm} thickness.
Based on the Beer-Lambert law~\cite{May2020} $I=I_0 \, \,T\left( \lambda \right)$, where $I$ denotes the measured intensity, $I_0$ the intensity of the light source, and $T\left(\lambda\right) = \mathrm{exp}\left(-\alpha \left( \lambda \right) \,\, d \right)$ the transmission with $\alpha\left( \lambda \right)$ representing the wavelength dependent absorption coefficient and $d$ the medium thickness, the transmission can be determined.
Rearranging results in $T_{10}\left( \lambda \right) = I_{30}/I_{20}$, with the subscripts denoting the membrane thicknesses in nm, assuming $I_0$ being identical.
Fig.~\ref{fig:rescaledSpectra}a) shows the obtained transmission in the range of the two \ch{He2}$^*$ excimer continua compared to the literature values of \ch{Si3N4} with the densities in question and \ch{SiO2} as oxidation of the membrane was visible in the EDX analysis.
For wavelength above \SI{70}{nm} the values fit to the predicted values of \ch{Si3N4} with the typical density of \SI{3.44}{\g\per\cubic\cm}.
Below \SI{70}{nm} the transmission increases and surpasses even the literature prediction for a density of \SI{2.94}{\g\per\cubic\cm}.
It is expected, that the experimentally determined values in this wavelength region are falsified by intensity changes of the VUV radiation source in between the measurements.
The measured intensities in this region also showed the strongest fluctuations and changes during the repeated measurements, therefore, it is expected to be most sensitive to changes in the plasma caused by, e.g., changing gas purity or temperature changes of the electrodes.
Overall, the values obtained in the region of the second \ch{He2}$^*$ excimer continuum (above \SI{70}{nm}) confirm the literature values and that in terms of transmission the typical density of \SI{3.44}{\g\per\cubic\cm} can be assumed for these \ch{Si3N4} membranes despite the indication of oxidation.

To visualize the influence of the \ch{Si3N4} membrane on the obtained spectra the measured intensities for the evacuated monochromator were divided by their corresponding wavelength dependent transmission values from literature~\cite{Hen1993,CXRO2024} for a thickness of \SI{20}{nm} and a density of \SI{3.44}{\g\per\cubic\cm} to calculate the intensity before transmission through the membrane for the wavelength region up to \SI{119}{nm}, as this wavelength marks the onset of second diffraction order of the first continuum.
Therefore, it is not fully clear from which radiation wavelength the measured intensity origins and the region above \SI{119}{nm} will be neglected.
Fig.~\ref{fig:rescaledSpectra}b) shows the original as well as the rescaled spectra normalized to the first \ch{He2}$^*$ excimer continuum.
The first \ch{He2}$^*$ excimer continuum is still the most intense feature of the spectrum unaffected by the transmission function of the \ch{Si3N4} membrane.
This is, as already mentioned, in contrast to previously reported VUV spectra~\cite{Gol2020,Liu2020,Sgo2024} and better corresponds to spectra obtained at slightly reduced pressures within differential pumping systems~\cite{Kur2001}.
The influence of the \ch{Si3N4} membrane increases to larger wavelength as given by the transmission curve~\cite{Hen1993,CXRO2024}, but overall is lower compared to the influence of alternative methods on atmospheric pressure plasmas~\cite{Gol2020,Liu2020,Sgo2024}.

Summarizing, it could be shown that the assumption of \ch{He} being transparent for VUV photons with wavelengths above \SI{50.48}{nm} is not justified for \ch{He} environments with pressures above \SI{5e4}{Pa} as photo-induced excimer formation results in absorption at least in the \SIrange{59}{61.8}{nm} region, probably even at all wavelength up to \SI{61.8}{nm}. Utilizing a \SI{20}{nm} thin \ch{Si3N4} membrane as entrance window for VUV photons produced at atmospheric pressure is an excellent alternative to common window materials not existing for wavelength below \SI{110}{nm} or other approaches utilizing aerodynamic windows or differential pumping systems. This was demonstrated using an atmospheric pressure plasma jet as VUV photon source.

\begin{acknowledgments}
The authors would like to thank Hendrik Kersten for fruitful discussions and Volker Rohwer for technical assistance in the lab. Funding of the German Research Foundation (DFG, Project Number 561666288, Grant Number HA 10873/1-1) is gratefully acknowledged. Further, L. H. gratefully acknowledges funding from the Postdoc Center of Kiel University via a CAU start grant.
\end{acknowledgments}

~\\
The authors have no conflicts to disclose.

~\\
\textbf{Luka Hansen:} conceptualization, formal analysis, funding acquisition, methodology, resources, supervision, visualization, writing/original draft preparation, writing/review \& editing. \textbf{Görkem Bilgin:} formal analysis, investigation, writing/review \& editing. \textbf{Jan Benedikt:} conceptualization, funding acquisition, methodology, resources, supervision, writing/review \& editing. 

~\\
The data that support the findings of this study are available from the corresponding author upon reasonable request.

\bibliography{Main.bib}

\end{document}